\documentclass[conference]{IEEEtran}
\IEEEoverridecommandlockouts
\usepackage{cite}
\usepackage{amsmath,amssymb,amsfonts}
\usepackage{algorithmic}
\usepackage{graphicx}
\usepackage{textcomp}
\usepackage{xcolor}

\def\BibTeX{{\rm B\kern-.05em{\sc i\kern-.025em b}\kern-.08em
    T\kern-.1667em\lower.7ex\hbox{E}\kern-.125emX}}
    
\usepackage{booktabs} 
\usepackage{multirow}
\usepackage{booktabs}
\usepackage{graphicx}  
\usepackage{diagbox}
\usepackage[braket]{qcircuit} 
\usepackage{threeparttable} 
\usepackage{stfloats} 
\usepackage{enumitem} 
\usepackage{fancyheadings} 
\usepackage{wrapfig} 

\usepackage{times}
\usepackage{graphicx}
\usepackage{epsf}
\usepackage{verbatim}
\usepackage{url}
\usepackage{color}
\usepackage{alltt}

\newcommand{\Caption}{\caption}

\newcommand{\CodeIn}[1]{{\small\texttt{#1}}}

\newcommand{\Comment}[1]{}

\newenvironment{CodeOut}{\begin{scriptsize}}{\end{scriptsize}}

\newcommand{\SmallSpace}{\vspace*{-1.4ex}}

\begin{document}

\title{Bugs4Q: A Benchmark of Real Bugs for Quantum Programs
}

\author{\IEEEauthorblockN{Pengzhan Zhao}
\IEEEauthorblockA{\textit{Kyushu University}
}
\and
\IEEEauthorblockN{Jianjun Zhao}
\IEEEauthorblockA{\textit{Kyushu University}
}
\and
\IEEEauthorblockN{Zhongtao Miao}
\IEEEauthorblockA{\textit{Kyushu University}
}
\and
\IEEEauthorblockN{Shuhan Lan}
\IEEEauthorblockA{\textit{Kyushu University}
}

}

\maketitle

\begin{abstract}
Realistic benchmarks of reproducible bugs and fixes are vital to good experimental evaluation of debugging and testing approaches. However,  there is no suitable benchmark suite that can systematically evaluate the debugging and testing methods of quantum programs until now. This paper proposes Bugs4Q, a benchmark of thirty-six real, manually validated Qiskit bugs from four popular Qiskit elements (\texttt{Terra}, \texttt{Aer}, \texttt{Ignis}, and \texttt{Aqua}), supplemented with the test cases for reproducing buggy behaviors. Bugs4Q also provides interfaces for accessing the buggy and fixed versions of the Qiskit programs and executing the corresponding test cases, facilitating the reproducible empirical studies and comparisons of Qiskit program debugging and testing tools. Bugs4Q is publicly available at \texttt{https://github.com/Z-928/Bugs4Q}.
\end{abstract}


\section{Introduction}

Software bugs have a significant impact on the economy, security, and quality of life. 
An appropriate method of bug finding can quickly help developers locate and fix bugs. Many software engineering tasks, such as program analysis, debugging, and software testing, are dedicated to developing techniques and tools to find and fix bugs. 
In general, these techniques and tools should be evaluated on real-world, up-to-date bug benchmark suites so that potential users can know how well they work. Such a benchmark suite should contain fail-pass pairs, consisting of a failed version, including a test set that exposes failures, and a passed version, which includes changes that fix failures. Based on this, researchers can evaluate the effectiveness of techniques and tools for performing bug detection, localization, or repair. As a result, research progress in this field is closely dependent on high-quality bug benchmark suites.

Quantum programming is the process of designing and constructing executable quantum programs to achieve a specific computational result. Several quantum programming approaches are available recently to write quantum programs, for instance, Qiskit~\cite{ibm2021qiskit}, Q\#~\cite{svore2018q}, and Scaffold~\cite{abhari2012scaffold}. The current research in quantum programming focuses mainly on problem analysis, language design, and implementation. Despite their importance, program debugging and software testing have received little attention in the quantum programming paradigm. The specific features of superposition, entanglement, and no-cloning introduced in quantum programming, make it difficult to find bugs in quantum programs~\cite{miranskyy2020your}. Several approaches have been proposed for debugging and testing quantum software~\cite{li2020projection,huang2019statistical,honarvar2020property,miranskyy2019testing,ali2021assessing} recently, but the debugging and testing remain challenging issues for quantum software~\cite{zhao2020quantum,miranskyy2019testing}.

Researches on bug benchmark suites for classical software have been studied extensively~\cite{do2005supporting,dallmeier2007extraction,just2014defects4j,lu2005bugbench,gyimesi2019bugsjs,hutchins1994experiments,le2015manybugs}, but few for quantum software. Recently, Campos and Souto~\cite{campos2021qbugs} proposed some initial ideas on building a bug benchmark for quantum software debugging and testing experiments.

We may not know which debugging or testing methods are suitable for quantum software without a suitable bug benchmark suite for evaluating these tools, and this may pose some restrictions on the research and development of quantum software debugging and testing techniques. 
As the first step towards evaluating quantum software debugging and testing tools, this paper proposes Bugs4Q, a benchmark of thirty-six real, manually validated Qiskit bugs from four popular Qiskit elements (\texttt{Terra}, \texttt{Aer}, \texttt{Ignis}, and \texttt{Aqua}), supplemented with the test cases for reproducing buggy behaviors. Bugs4Q has made the following contributions:

\begin{itemize}[leftmargin=2em]
\setlength{\itemsep}{3pt}
  \item Bugs4Q collects reproducible bugs in Qiskit programs and supports downloading and running test cases for quantum software testing. Each actual bug and the corresponding fixes are publicly available for research.
  \item Bugs4Q collects almost all the existing bugs of Qiskit on GitHub and updates them in real-time, including the four elements of \texttt{Terra}, \texttt{Aer}, \texttt{Ignis}, and \texttt{Aqua}.
  Furthermore, these programs are sorted separately and filtered except for the bugs with originally available test cases and support for reproduction.
  \item Bugs4Q provides a database that includes an analysis of bug types to classify existing bugs for experimental evaluation of isolated bugs.
\end{itemize}

The rest of the paper is organized as follows. Section~\ref{sec:bug-database} briefly describes Bugs4Q, a bug benchmark suite for Qiskit. Section~\ref{sec:Reproduction} introduces the process of manually reproducing bugs. 
Section~\ref{sec:Common types of bug} describes the specific types of bugs in Qiskit with examples. 
Related work is discussed in Section~\ref{sec:related-work}, and concluding remarks are given in Section~\ref{sec:conclusion}.

\section{Bugs4Q Benchmark Suite}
\label{sec:bug-database}
To make sure that we can build a benchmark of real bugs, we have collected the existing bugs in the version control history and the real fixes provided by the developers.

Table~\ref{table:benchmark} shows all programs and the numbers of corresponding real bugs that are available in the bug database of Bugs4Q.
In order to achieve benchmark rigor, each real bug must have its original bug version as well as a fixed version. This requires us to extract the relevant description of the bug and refer to its fixed commit. Moreover, the bugs we collect must comply with the following requirements:

\begin{itemize}[leftmargin=2em]
\setlength{\itemsep}{3pt}
  \item \textbf{Related to source code.} The reason for the bug is on the source files of the build system, but not the test files or the underlying files that build the Qiskit programs.
  
  \item \textbf{Related to quantum program.} The bug should have an impact on the operation or the outcome of the quantum program. Problems caused by configuration files required to run quantum programs or classical parts of quantum algorithms are not included in our database.
  
  \item \textbf{Reproducible.} More than one test case must be used to demonstrate the bug, and the bug must be reproducible under certain requirements. Depending on the nature of the quantum program, for example, the presence of probabilistic output causes the program not to be able to reproduce the results completely. It can lead to bugs that are difficult to reproduce in a controlled environment.
  
  \item \textbf{Isolated.} Fixes submitted by developers should also be related to the source files. Irrelevant changes need to be removed, and there is no code refactoring due to version changes. Excessive source file changes that are too complex will be incorporated into our database repository later after careful verification of isolation.
\end{itemize}

\begin{table}[t]
\caption{\label{table:benchmark} Programs and number of real bugs available in the initial version of Bugs4Q}
\scriptsize{
\begin{center}
\renewcommand\arraystretch{1.0}
\begin{tabular} {|l|l|r|r|r|r|}
\hline 
\textbf{Program}&\textbf{Source}&\textbf{Bugs}&\textbf{KLOC}&\textbf{Test KLOC}&\textbf{Tests}\\
\hline

\hline Terra & IBM Qiskit & 27 & 139 & 56 &467\\
\hline Aer & IBM Qiskit & 5 &  62& 18 &149\\
\hline Ignis & IBM Qiskit & 1 & 11 & 3 &59\\
\hline Auqa & IBM Qiskit & 3 &  56 & 17 &211\\
\hline

\end{tabular}
\end{center}
}
\vspace*{-7ex}
\end{table}

\begin{figure}[h]
    \centering
    \includegraphics[width=6.5cm]{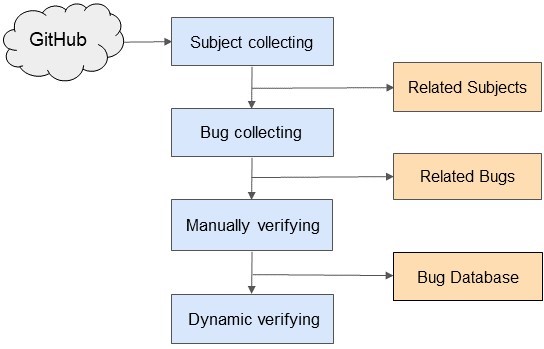}
    \caption{The building process of the benchmark database}
    \label{fig:galaxy}
    \vspace*{-2mm}
\end{figure}

Figure~\ref{fig:galaxy} depicts the main process of building our benchmark.
First, we look for programs on  as criteria for our base database classification. After that, we collect the issues with \textit{bug} tags and incorporate them into our first version of the database for manual validation.
We then manually reproduce the bugs for further filtering and place those that meet the isolation criteria into our version-2.0 database for dynamic validation. Finally, extraneous patches will be cleaned up to complete the final version of the benchmark database.


\subsection{Selecting Subject Programs}
For project program selection, we only target Qiskit programs that are relatively well used on GitHub. We use the GitHub's \textit{issue} tab to find bugs in the program and collect both the \textit{buggy} version and the \textit{fixed} version. 
We have manually collected all the issues with \textit{bug} tags.
Our study is based on four elements of \texttt{Terra}, \texttt{Aer}, \texttt{Ignis}, and \texttt{Aqua} in Qiskit. 
A brief descriptions of the four elements, which comes from the official description of Qiskit~\cite{ibm2021qiskit}, are given as follows:
 
\begin{itemize}[leftmargin=2em]
\setlength{\itemsep}{1pt}
  \item \texttt{Terra} is the foundation on which the rest of Qiskit is built. 
  \item \texttt{Aer} provides high-performance quantum computing simulators with realistic noise models.
  \item \texttt{Ignis} provides tools for quantum hardware verification, noise characterization, and error correction. 
  \item \texttt{Aqua} provides algorithms for quantum computing applications.
\end{itemize}

In this process, we will choose quantum-related bugs and exclude pure classical program bugs. Quantum programs in general are operations on qubits, so bugs in quantum programs are defined as bugs that occur when operating on qubits. A complete quantum program includes initialization of qubits, gate operations, and measurement operations, as well as deallocations. As shown in Figure~\ref{fig:mesh10} is a running example of a quantum program.
For example, if a quantum gate (e.g. \CodeIn{circuit.cx(0,1)}) in a Qiskit program does not achieve a quantum state transition or a measurement result is wrong, it will be classified as a Qiskit bug.

\begin{figure}[t]
  \begin{CodeOut}
\footnotesize{
  \begin{alltt}
    simulator = \textbf{Aer.get_backend}('qasm_simulator')

    qreg = \textbf{QuantumRegister}(3)
    creg = \textbf{ClassicalRegister}(3)
    circuit = \textbf{QuantumCircuit}(qreg, creg)

    circuit.\textbf{h}(0)
    circuit.\textbf{h}(2)
    circuit.\textbf{cx}(0, 1)
    circuit.\textbf{measure}([0,1,2], [0,1,2])
    job = \textbf{execute}(circuit, simulator, shots=1000)
    result = job.\textbf{result}()
    counts = result.\textbf{get_counts}(circuit)
    \textbf{print}(counts)
  \end{alltt}
}
\end{CodeOut}
\Caption{\label{fig:mesh10}A simple quantum program in Qiskit}
\end{figure}

\subsection{Collecting Bugs}
Since there are not particularly many bugs in Qiskit, we collect them all manually.
The bug collection process consists of two steps.

\subsubsection{Collecting and Filtering Bugs}
For each element in Qiskit, we focus on collecting issues from \textit{closed} tags on GitHub. We also collect bugs with \textit{open} status and mark them so that they can be put into our database as soon as they are submitted for fixing. 
For bugs in \textit{open} status, we temporarily store them in a branch but leave them unprocessed. For bugs in \textit{closed} status, we collect bugs and submitted fixes according to their IDs (e.g., \#1324). Of course, in our benchmark library, we use our own bug ID as the index.

In this work, we only consider bugs that have been fixed and are fully reproducible. So the bugs which have been reported but are in open status in GitHub issues and bugs that are not related to quantum programs will be filtered out. We also discard the case of having multiple fixes for bugs, i.e., having multiple fix links. Besides, bugs that disappear due to version changes are also not considered. After this work, we built the original bug database, which contains 206 quantum program-related bugs. As more and more bugs are raised, our database will be updated as we go forward.

\subsubsection{Screening Test Cases}
In order to reproduce the bugs more accurately, we choose the original program proposed by the developer in the bug report as our test case. The program proposed by the developer will be collected and tagged with the test in our library, accompanied by the second screening. There are 115 bugs with original test cases. For the bugs, without test cases, we will write the corresponding test cases of the bugs in our future work. 

\subsection{Manually Verifying and Reproducing Bugs}
We manually check that our requirements are met for each bug that has a test case. All bugs that can be reproduced and meet the isolation criteria will be placed in our final benchmark suite. For incomplete test cases, we modify the recovery procedure as much as possible to achieve its proper operation. Nevertheless, in addition to errors in the original test cases, it is not uncommon to have too many and complex revision submissions that are not reproducible. As in \texttt{Ignis}, many files have the suffix \texttt{hpp} and \texttt{cpp} instead of \texttt{py}. We chose to forgo collecting them into the current version of the benchmark suite. On the other hand, some of the submitted fixes have no impact on the bug recovery, so we only keep the fixes that impact the bug.
Table~\ref{table:4} shows the number of bugs with test cases that can not be reproduced as well as the reasons. The results of the manual validation show that only thirty-six bugs have been successfully reproduced and isolated. The process of reproduction is still ongoing.

\begin{table}[t]
\caption{\label{table:4} Manual Validation Reports}
\footnotesize{
\begin{center}
\renewcommand\arraystretch{1.0}
\begin{tabular}{|p{6.5cm}|r|}
\hline 
\textbf{Reasons of Non-reproducible Bugs}&\textbf{Number}\\\hline

\hline One bug corresponds to multiple fixes &  6\\
 Fix too many files&  2\\
 Fixes not work&  9\\
 Actual error does not match the submitted bug&  2\\
 Program reports an error and cannot run& 35\\
 Other& 9\\
\hline 
\textbf{Final Bug Database}&\\
\hline Bugs with original test cases & 36\\
\hline
\end{tabular}
\end{center}
}
\vspace*{-7mm}
\end{table}

\subsection{Sanity Checking through Dynamic Validation}
In order to better reproduce the bug, we try to implement an automated approach. We first implement version control, calling the corresponding version of Qiksit for different bugs. After calling the bug indexed by ID, the file of the bug version will replace the corresponding file in Qiskit. Finally, the same process is implemented for the fixed version. As the test suite continues to improve, the version control environment will be ported to more platforms.

\begin{table}[h]
\caption{\label{table:rules} Restrictions on reproducing bugs}
\footnotesize{
\begin{center}
\renewcommand\arraystretch{1.0}
\begin{tabular}{|l|p{6cm}|}
\hline \textbf{Restrictions} &\textbf{Description}\\\hline

\hline Isolation & Each fix submission can only address one bug, and that bug cannot exist on top of any other bug\\
\hline Complication & One bug corresponds to multiple fixes submitted, and it is not possible to identify the specific part of the source file that was modified\\
\hline Reconfiguration & Fixed commits are file rewrites caused by refactoring or version changes \\
\hline Dependencies & The fixed commit introduces a new library\\
\hline

\end{tabular}
\end{center}
}
\end{table}

\section{Available Bugs Reproduced}
\label{sec:Reproduction}
This section describes the process of manually reproducing bugs. 
Among the bugs we reproduced, most of them are programs that use the Qiskit simulators. Therefore most of the recurrence process is implemented manually on our PC side. If there is a program connected to the IBM cloud backend, we also reproduce these operations as much as possible.
Bugs are complex to reproduce, so first, they need to meet some rules. Table~\ref{table:rules} shows the restrictions on reproducing bugs. We separate each bug, clean up irrelevant changes in advance, ignore some description files, and keep only the original files related to the bug and the fixed files. Any bugs or fixes with the above characteristics will be filtered out. 

\subsection{Restoring Version Environment} 
\begin{figure}[t]
    \centering
    \includegraphics[width=6cm]{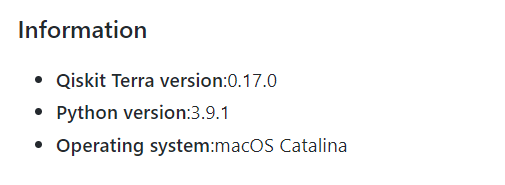}
    \caption{Version information of bug submission}
    \label{fig:2}
\end{figure}

\begin{figure}[htp]
    \centering
    \includegraphics[width=8.5cm]{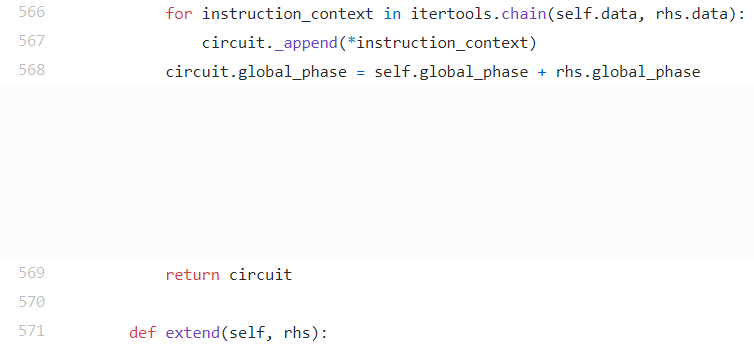}
    \caption{The partial code of buggy file}
    \label{fig:3}
\end{figure}

We configure the environment based on the version information submitted by the bug raiser in Figure~\ref{fig:2}.
This is error \#5908 \footnote[1]{https://github.com/Qiskit/qiskit-terra/issues/5908} as an example, its proposed version is \CodeIn{terra 0.17.0}.
After that, we find the file location and the repaired file code in a fixed commit. Then we manually restore the fixed code to the code at the time of the buggy. Finally, we replace the files in the environment with the restored source files. 
In this sample, the blue code section in Figure~\ref{fig:4} is the code added by the repaired file. We restore it to the buggy state as represented in Figure~\ref{fig:3}.

\subsection{Running Test Cases} 
\begin{figure}[t]
    \centering
    \includegraphics[width=8.5cm]{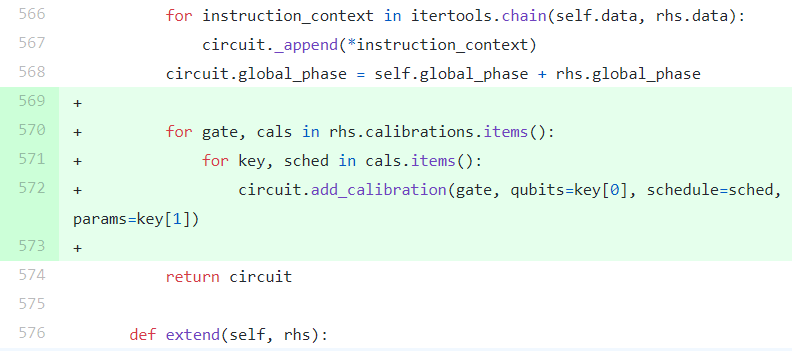}
    \caption{The partial code of fixed file}
    \label{fig:4}
    \vspace*{-4ex}
\end{figure}

We first select the program code provided by the bug raiser. 
Figure~\ref{fig:5} shows the program code in the error submission message that does not meet the program run criteria.
The code will be downloaded, fixed, and placed in our database as original test cases for verification. When we run the program in the configured environment, the result is consistent with the description of the bug submission message, which proves that the bug has been successfully reproduced. Next, the fixed file represented in Figure~\ref{fig:4} replaces the file in the environment. If the bug disappears, the program runs successfully and is consistent with the description of the fix, the test passes.

\begin{figure}[t]
  \begin{CodeOut}
\scriptsize{
  \begin{alltt}
  
  
qc = \textbf{QuantumCircuit}(2)
qc.\textbf{rzx}(0.1,0,1)
    
pass_ = \textbf{TemplateOptimization}(**rzx_templates(['zz2']))
qc_cal = \textbf{PassManager}(pass_).run(qc)
pass_ = \textbf{RZXCalibrationBuilder}(backend)
qc_cal = \textbf{PassManager}(pass_).run(qc_cal)
    
qc2 = \textbf{QuantumCircuit}(2)
qc2 += qc_cal
    
\textbf{print}(schedule(transpile(qc,backend),backend).duration)
\textbf{print}(schedule(transpile(qc_cal,backend),backend).duration)
\textbf{print}(schedule(transpile(qc2,backend),backend).duration)
  \end{alltt}
}
\end{CodeOut}
\Caption{\label{fig:5} The code in the problem description}
\vspace*{-4mm}
\end{figure}

\subsection{Filter Information}
Any commit fix has a \textit{Files changes} section. We can sort through the modified files and code segments to locate the fixes that only affect the bug.
When multiple buggy files exist in the process of reproducing bugs, we will recover them one by one because there are some files whose restoration does not affect the bug.
\begin{wrapfigure}{r}{4cm}
    \centering
    \includegraphics[width=4cm]{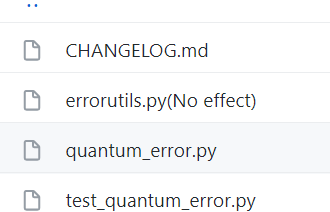}
    \caption{Fixed files of issue \#580}
    \label{fig:6}
\end{wrapfigure}
Figure~\ref{fig:6} shows the four files which have been used to fix the error\#580~\footnote[2]{https://github.com/Qiskit/qiskit-aer/issues/580} in \CodeIn{Aer}. The only two of these files, \CodeIn{quantum\_error.py} and \CodeIn{errorutils.py} are the source files used to fix the bugs. Once sift out the other two files, we can reproduce the bug with the above two source files in turn. We find that only the file \CodeIn{quantum\_error.py} affects the bug. Thus, the file \CodeIn{errorutils.py} is eliminated.

\subsection{Representation in Benchmark Suite} 

The reproduced available bugs will be added to our database.
As shown in Table~\ref{table:3}, \textit{Buggy} is the source file that caused the bug before the fix. \textit{Fixed} is the file after the fix for the bug. \textit{Test} is the original test case.
In order to be able to document each bug in detail, we provide a detailed description of each bug and links to local files of our organization for them. Furthermore, \textit{Issue No} and \textit{Modify} are linked to bugs and fixes committed in GitHub, respectively.

\begin{table*}[t]
\caption{\label{table:3} Example of a benchmark database for Bugs4Q}
\footnotesize{
\begin{center}
\renewcommand\arraystretch{1.0}
\begin{tabular}{|l|l|l|l|l|l|l|l|l|l|l|}
\hline 
\textbf{Bug ID}&
\textbf{Issue No}&
\textbf{Buggy}&
\textbf{Fixed}&
\textbf{Modify}&
\textbf{Status}&
\textbf{Version}&
\textbf{Type}&
\textbf{Test}&
\textbf{Issue Registered}&
\textbf{Issue Resolved}\\
\hline
\hline 6 & \#5914 & Buggy & Fixed & Mod & Resolved	& 0.16.4 & Bug & Test & Feb 27, 2021 & Feb 28, 2021\\
\hline 7& \#5908 & Buggy & Fixed & Mod & Resolved	& 0.17.0 & Bug & Test & Feb 26, 2021 & Mar 1, 2021\\

\hline
\end{tabular}
\end{center}
}
\vspace*{-4mm}
\end{table*}

\section{Analysis}
\label{sec:Common types of bug}
After reproducing each bug and populating the benchmark database, we also summarize some bug types as shown in Table~\ref{table:5}.
To prove that the reproduced bug is consistent with the original description, we need to compare the results of the program run with the submitted bug information. For example, whether the error output we reproduced is consistent with the result in the description, or whether the program we reproduced reports errors are consistent with the submitted error messages. In this process we summarize the bugs we encountered.
\begin{table}[h]
\caption{\label{table:5} Classification of Bugs}
\footnotesize{
\begin{center}
\renewcommand\arraystretch{1.0}
\begin{tabular}{|p{6cm}|r|}
\hline 
\textbf{Bug Type}&\textbf{Number}\\\hline

\hline Output wrong &  15\\
\hline Noise simulation error& 1 \\
\hline Lost information&  3\\
\hline Throwing exceptions&  13\\
\hline Circuit diagram drawing error& 4\\
\hline
\end{tabular}
\end{center}
}
\end{table}

\subsection{Output Wrong}
Output errors are the bugs we are most concerned about, which are not easily discovered but play a critical role in quantum programs. Erroneous output results can mislead program users. Here we reproduce a simple example in \texttt{Aqua}, with the issue number of \#1324 \footnote[3]{https://github.com/Qiskit/qiskit-aqua/issues/1324}.
\begin{figure}[t]
  \begin{CodeOut}
\footnotesize{
  \begin{alltt}
  
  
    Output before fixed:  v1 = 32.0   v2 = 8.0
    Output after fixed:   v1 = 8.0    v2 = 8.0
  \end{alltt}
}
\end{CodeOut}
\Caption{\label{fig:mesh1}An example of output wrong}
\end{figure}

Considering the code snippet in Figure~\ref{fig:mesh1}. The program that calls the \CodeIn{CircuitSampler} method and finds that \CodeIn{v1} and \CodeIn{v2} should have output the same result. But they are only the same when \CodeIn{coeff=1}, otherwise they often have different results.
The file to fix this bug is \CodeIn{vector\_state\_fn.py} only, which is the source file in \CodeIn{qikist/aqua}.
Such bugs account for a large proportion of the bugs we reproduce.

\subsection{Noise Simulation Error}
Due to the inherent nature of quantum computer hardware, the presence of noise makes programs that actually run in quantum computers not extraordinarily stable.
So Qiskit provides \texttt{Aer} which can be implemented to simulate noise on a classical computer. This provides great convenience for us to study real quantum programs. Therefore it is particularly important to target the bugs of quantum noise simulators.

\begin{figure}[t]
  \begin{CodeOut}
\footnotesize{
  \begin{alltt}
Output before fixed:   
        From config : ['id', 'rz', 'sx', 'x', 'cx']
        From the noise model : 
                      ['cx', 'id', 'sx', 'u3', 'x']
                 
Output after fixed:
        From config : ['id', 'rz', 'sx', 'x', 'cx']
        From the noise model : 
                      ['id', 'rz', 'sx', 'x', 'cx']
  \end{alltt}
}
\end{CodeOut}
\Caption{\label{fig:mesh2}An example of noise simulation error }
\end{figure}

As shown in Figure~\ref{fig:mesh2}, which is a program that imports the base gate from the quantum simulation backend from issue \#1107\footnote[4]{https://github.com/Qiskit/qiskit-aer/issues/1107}. By default, the noise model contains in its usual base gate the \CodeIn{id} and \CodeIn{U3}.
The purpose is that circuits could be executed even if the developer did not define noise on all gates.
However, a bug in running the program due to a change in the default base gate of the IBM Q backend prevents it from working correctly. 
The bug here is that the \CodeIn{u3} gate should not appear in the noise model, but rather the \CodeIn{X} gate.
The fix for this commit is that the noise model will always have the same base gate as the backend base gate, regardless of whether the instruction has an error in the noise model.
This type of error is only found in the \CodeIn{Aer} element and is not common.

\subsection{Lost Information}
Lost information, i.e., the program does not implement a specific function.
Terra is the most used element and the one with the most bugs filed. There are many times the fix for bugs in other elements of the commit is to modify the files in the \texttt{terra}.
The sample we cite is \#5908, as shown in Figure~\ref{fig:5}.
This is a program that combines a conventional \CodeIn{QuantumCircuit} with a calibration circuit. The buggy version of this program uses the default gate circuit, and the calibration circuit information is missing.
The error output with \CodeIn{name = "sched4"} indicates that \CodeIn{QuantumCircuit}'s \CodeIn{+=} does not remain calibrated, thus causing the problem.
Such a bug is not closely related to traditional quantum circuits and is not common.

\subsection{Throwing Exceptions}
Throwing an exception is as common and basic as an output error. Program errors and output errors account for almost all of the bugs in our Database. For example,  Figure~\ref{fig:mesh4}, which is from \CodeIn{Qiskit Terra}, \#2369\footnote[5]{https://github.com/Qiskit/qiskit-terra/issues/2369}, is a simple program of using indexes and bits as parameters. Until it is fixed, this bug can be considered as a bug pattern. However, this issue is fixed as a bug here. So we collected it into our Database. In general, this kind of bug can be understood as the parameters of the called method do not support string types, or more precisely, only integer types. The problem here is that the gate parameter could not support a mix of indexes and bits. A commit fix made it possible for the gate parameter to support them.

\begin{figure}[t]
  \begin{CodeOut}
\footnotesize{
  \begin{alltt}
    Output before fixed: 
            QiskitError: "Type error handeling 
            [(QuantumRegister(1, 'q1'), 0), 1] 
            (<class 'list'>)"
            
    Output after fixed:
            Qubit ordering: 
            [Qubit(QuantumRegister(1, 'q2'), 0), 
             Qubit(QuantumRegister(1, 'q1'), 0)]
            Classical bit ordering: 
            [Clbit(ClassicalRegister(2, 'c'), 0), 
             Clbit(ClassicalRegister(2, 'c'), 1)]
  \end{alltt}
}
\end{CodeOut}
\Caption{\label{fig:mesh4}An example of throwing exceptions}
\end{figure}

In addition to the error types described above, we constantly summarize other bug types, such as quantum circuit diagram drawing errors.
Qiskit programs generally generate circuit diagrams, which serve only to represent the process of changing quantum states and do not impact on the program's execution. However, wrong circuit diagrams can also mislead users, and therefore we have collected them into one type.
In our future work, we will add the types of bugs to our benchmark and propose a database with bug types as classification criteria for better use by researchers and developers.

\section{Related Work}
\label{sec:related-work}

Many benchmark suites have been proposed to evaluate debugging and testing methods for classical software. 
The Siemens test suite~\cite{hutchins1994experiments} is one of the first bug benchmark suites used in testing research. It consists of seven C programs, which contain manually seeded faults. 
The first widely used benchmark suite of real bugs and fixes is SIR (Software Artifact Infrastructure Repository)~\cite{do2005supporting}, which enables reproducibility in software testing research. SIR contains multiple versions of Java, C, C++, and C\# programs which consist of test suites, bug data, and scripts. 
Other benchmark suites include Defects4J~\cite{just2014defects4j} and iBug~\cite{dallmeier2007extraction} for Java, BenchBug~\cite{lu2005bugbench}, ManyBug (and InterClass)~\cite{le2015manybugs}, and B{\footnotesize UGS}JS~\cite{gyimesi2019bugsjs} for JavaScirpt projects.

However, all the benchmarks mentioned above focus on the classical software systems, and therefore cannot be used for evaluating and comparing quantum software debugging and testing methods. 

Perhaps, the most related work with ours is QBugs proposed by Campos and Souto~\cite{campos2021qbugs}, which is a collection of reproducible bugs in quantum algorithms for supporting controlled experiments for quantum software debugging and testing. QBugs proposes some initial ideas on building a benchmark for an experimental infrastructure to support the evaluation and comparison of new research and the reproducibility of published research results on quantum software engineering. It also discusses some challenges and opportunities on the development of QBugs. However, at the time of writing this paper, the benchmark proposed by Campos and Souto~\cite{campos2021qbugs} was not available to assess its details and usability. Our Bugs4Q, on the other hand, aims to construct a bug benchmark suite of real bugs derived from four real-world IBM Qiskit programs for quantum software debugging and testing, with real-world test cases for reproducing the buggy behaviors of identified bugs.  

\section{Concluding Remarks}
\label{sec:conclusion}
As quantum computers gradually come into the limelight, research into quantum programs has intensified, with analysis and testing techniques for quantum programs becoming an important part of the process.
This paper proposes Bugs4Q, a benchmark of thirty-six real, manually validated Qiskit bugs from four popular Qiskit programs, supplemented with the test cases for reproducing buggy behaviors. Bugs4Q also provides interfaces for accessing the buggy and fixed versions of the Qiskit programs and executing the corresponding test cases, facilitating the reproducible empirical studies and comparisons of Qiskit analysis and testing tools.

In our future work, we would like to keep updating the Bugs4Q benchmark based on bug reports submitted on GitHub for Qiskit and continue to improve the tests to reproduce more bugs in Qiskit.
Our benchmark will be continuously maintained on an ongoing basis. As Qiskit versions are updated, we will propose new extensions while keeping the previous version and subsequently develop new frameworks.
Although quantum programs are now basically run with Qiskit simulators and the programs are small in size. However, we will try to replicate this on the QPU in a subsequent attempt to find out what kind of bugs occur in programs running on quantum computers.
We would also like to summarize more bug types for the commonality of bugs to be more easily uncovered when Bugs4Q is extended to other quantum programming languages.

\bibliographystyle{IEEEtran}
\bibliography{IEEEabrv,ase2021-nier}

\end{document}